\documentclass[12pt]{article}

\usepackage{amsfonts}
\usepackage{epsfig}
\usepackage{latexsym}
\usepackage{graphicx}
\def\bequ{\begin{equation}}
\def\eequ{\end{equation}}
\def\barr{\begin{array}}
\def\earr{\end{array}}

\def\ben{\begin{equation}}
\def\een{\end{equation}}
\def\bena{\begin{eqnarray}}
\def\eena{\end{eqnarray}}

\def\nn{\nonumber}
\newcommand{\sect}[1]{\setcounter{equation}{0}\section{#1}}
\renewcommand{\theequation}{\arabic{section}.\arabic{equation}}
\def\spa#1{\phantom{\fbox{\rule[-#1cm]{0cm}{0cm}}}}
\def\thefootnote{\fnsymbol{footnote}}
\renewcommand{\thefootnote}{\alph{footnote}}

\newcommand{\beqn}{\begin{equation}}
\newcommand{\eeqn}{\end{equation}}
\newcommand{\beqarr}{\begin{eqnarray}}
\newcommand{\eeqarr}{\end{eqnarray}}

\newcommand{\matc}{\begin{array}{c}}
\newcommand{\matcc}{\begin{array}{cc}}
\newcommand{\matccc}{\begin{array}{ccc}}
\newcommand{\matcccc}{\begin{array}{cccc}}
\newcommand{\emat}{\end{array}}

\begin{document}

\begin{titlepage}

August 2003         \hfill
\begin{center}

\vskip .90in
\renewcommand{\thefootnote}{\fnsymbol{footnote}}
{\large \bf Closed Geodesics on G\"odel-type Backgrounds}
\vskip .30in

Daniel Brace\footnote{email address: brace@physics.technion.ac.il}
\vskip .30in
{\it Department of Physics, Technion},
\\ {{\it Israel Institute of Technology}},
\\ {{\it Haifa 32000, Israel}}

\end{center}
\vskip 2.0in

\begin{abstract}
We consider radial oscillations of supertube probes in the G\"odel-type background which is U-dual
to the compactified pp-wave obtained from the Penrose limit of the NS five-brane near horizon 
geometry. The supertube probe computation can
be carried over directly to a string probe calculation on the U-dual background. The classical
equations of motion are solved explicitly. In general, the probe is not restricted to
travel unidirectionally through any global time coordinate. In particular, we find geodesics that close. 
\end{abstract}

\end{titlepage}

\newpage
\renewcommand{\thepage}{\arabic{page}}

\setcounter{page}{1}
\setcounter{footnote}{0}



\sect{Introduction}
It was shown by Gauntlett et al \cite{Gauntlett:2002nw} that low energy string 
theory admits supersymmetric solutions of the G\"odel type \cite{Godel:ga}. 
These homogeneous spaces have Closed Timelike Curves (CTCs) and Closed Null 
Curves (CNCs) which are homotopic to a point. It was then pointed out that 
these Closed Causal Curves (CCCs) are not present in dimensional 
upliftings of certain T-dual versions of these G\"odel spaces \cite{Herdeiro:2002ft}. 
The reason was understood in \cite{Boyda:2002ba}: the uplifted spacetime 
is a standard pp-wave of the type that has been recently studied as the Penrose 
limit of certain near horizon brane geometries \cite{Berenstein:2002jq}. 
Such realization raised the hope that string theory could shed light on 
one of the more important open problems in General Relativity, or more 
generally, in gravitational theories: are geometries with CCCs intrinsically 
inconsistent or is propagation of matter 
in geometries with CCCs intrinsically inconsistent? 

In this paper, we obtain results that suggest 
that string theory on certain supergravity backgrounds possessing CCCs may be problematic. By studying
the geodesics of $D2$ brane (supertube) probes on a certain G\"odel-type background, 
we are able to find classical trajectories that close on 
themselves in spacetime. We will show that after a finite
evolution in an affine parameter, an initial slice of D2 brane probe, which may be spacelike
and located within a causal region of the G\"odel space, 
may return to its original position and that further
evolution of its coordinate embedding is periodic in this affine parameter. 
We will refer to this type of solution of the probe equations of motion as a periodic geodesic.
A closed geodesic is then a periodic geodesic whose affine parameter has been periodically
identified so that the probe's Lorentzian worldvolume is finite. In the case of D-brane geodesics, there are extra worldvolume gauge fields that must share a common
period with the spacetime coordinates in order to make this identification. 
We will show that this is possible if a
certain combination of parameters is rational. In particular, this combination depends on
the parameter $f$ of the G\"odel background and on the radius of a compact direction, along with the
magnetic field and the momentum conjugate to the electric field on the brane. It is always possible
to choose the momentum conjugate to the electric field appropriately in order to obtain closed geodesics.
On the other hand, if this choice is not made or if the affine parameter is simply not
periodically identified, the D2-brane probe traveling along a periodic geodesic
leads to divergences in the energy momentum tensor. It is clear that
this renders the probe calculation inadequate, and the effects of gravitational backreation and self-interaction 
should  be considered. 
At the quantum level, the closed geodesics are potentially problematic. In analogy with point particle examples 
where closed one dimensional geodesics exist, one might expect pathologies to manifest themselves
in the full quantum theory. We leave these issues for future work. 
It is interesting to note that there are no closed timelike particle geodesics in these G\"odel-type Universes 
\cite{Brecher:2003rv},
so closed geodesics (with Lorentzian signature) 
only appear when considering stringy matter in these backgrounds.
Recent discussions of CCCs in string theory can be found 
in \cite{Boyda:2002ba, Harmark:2003ud, Dyson:2003zn, Biswas:2003ku, Drukker:2003sc, Hikida:2003yd, Brecher:2003rv, Herdeiro:2003un,BHH}.

The outline of the paper is as follows. In the following two sections, we solve the classical equations of
motion of the supertube probe, whose radius we allow to oscillate. In Section 4, by considering some
specific parameters, we are able to find probe geodesics that are periodic and discuss under which
conditions they can be closed. We briefly discuss the gravitational
coupling of the probes in Section 5. The details of the solution are worked out in Appendix A. In Appendix B,
we translate the supertube solutions into the U-dual string geodesics and make some brief comments about
string quantization on the compactified pp-wave background.

\sect{Supertube Probes}
The metric and background fields of the G\"odel Universe we will be studying are given 
by\footnote{In this paper, we will take $f$ to be positive.}
\bequ
\barr{c}
\displaystyle{ds^2=-\left[dt+fr^2d\theta\right]^2+dy^2+dr^2+r^2d\theta^2+\delta_{ij}dx^idx^j} \ , 
\spa{0.3} \\
\displaystyle{B_{NS}=fr^2dy\wedge d\theta \ , \ \ \ \ C^{(3)}=fr^2d\theta\wedge dt \wedge dy \ ,  \ \ \ \ C^{(1)}=-fr^2d\theta} \ , \earr \label{back} \eequ
which is a Type IIA supergravity background preserving one quarter of the maximal number of supersymmetries
\cite{Harmark:2003ud}. 
CCCs of the G\"odel background can be seen by considering the curve generated by 
$\frac{\partial}{ \partial \theta}$. This curve is spacelike for $r<f^{-1}$, null for $r=f^{-1}$, and 
timelike for $r>f^{-1}$. The surface defined by  $r=f^{-1}$ will be referred to as the velocity of light surface (VLS).
We will take the $y$ direction to be compact with period $2\pi L$.

In flat space, it was shown \cite{MT} that cylindrical $D2$ branes can be supported against collapse by angular momentum
generated by electric and magnetic fields on their worldvolume, and that these supertubes are
 $\frac{1}{4}$ BPS. More recently, it was shown \cite{Drukker:2003sc} 
that in the G\"odel background (\ref{back}) supertubes
continue to exist and preserve the same supersymmetry as the background itself.  

If we consider a cylindrical $D2$-brane extended in $y$ and 
wrapping the $\theta$ direction $N$ times, the system is 
described by a $U(N)$ gauge theory with twisted boundary conditions. 
Restricting attention to the $U(1)$ zero mode 
components of the field strength and radial mode, the
probe is described by a Born-Infeld Lagrangian with couplings to the background $RR$ fields. 
\bequ 
{\cal L} = -|N\, | \sqrt{-{\mbox d}{\mbox e}{\mbox t}(G + {\cal{F}})} -N\,  e^{\cal F} \sum C^{(n)} \, ,
\label{gen}
\eequ
where ${\cal F} = F - B_{NS}$. 
Using static gauge, we find that the Lagrangian for the supertube probe, centered at $r=0$ and extended in
the $\theta$ and $y$ directions of the background (\ref{back}), is given by\footnote{ Here,
we suppress various dimensionful parameters. We set $2\pi \alpha' = 1$, as well 
as $(2\pi)^{3/2}L = 1$. The latter can
be recovered by taking $N\rightarrow (2\pi)^{3/2}LN$.}
\bequ
{\cal{L}} = -|N| \sqrt{ (-{\dot{r}}^2 +\Delta^{-1})( {r^2}\Delta + \bar{B}^2 ) - \bar{E}^2 {r^2}\Delta } 
- Nfr^2  + Nfr^2E \, ,
\label{olag}
\eequ 
where
\bequ
\Delta = 1 -f^2 r^2 \ , \ \ 
\bar{B} = B - fr^2  \ , \ \
\bar{E} = E - { f\bar{B}}\Delta^{-1} \, ,
\eequ
and $E$ = $F_{0y}$ and $B$ are the electric and magnetic fields on the brane. 
We choose to consider only configurations\footnote{ Choosing temporal gauge (on the probe) $A_0 = 0$, one must
enforce the Gauss law constraint $\partial_\theta \Pi_\theta + \partial_y \Pi_y = 0$. It is consistent
to set $F_{0 \, \theta} = \Pi_\theta = 0$ for vanishing $\partial_y \Pi_y$. On the other hand, as opposed to the
flat space case, it is not
consistent to set $E = F_{0y} = 0$ since $E$ and $\Pi_y$ are not directly proportional and 
differ by $r$ dependent terms.} with $F_{0 \,\theta} = 0$. 
The Hamiltonian, ${\cal H} = (N\Pi) E + (|N |P_r) \, \dot{r} -{\cal L}$, can be obtained.
\bequ
{\cal{H}} = \frac{s \, |N|}{r\Delta} \sqrt{ {P_r}^2 r^2 \Delta + ( r^2 \Delta+\bar{\Pi}^2 )
(r^2 \Delta + \bar{B}^2) }
+\frac{Nf}{\Delta}(r^2\Delta + \bar{\Pi}\bar{B}) \, ,
\label{Ham}
\eequ
where
\bequ
P_r = |N|^{-1}\frac{\partial{\cal{L}}}{\partial \dot{r}} \ , \ \    \Pi = N^{-1}\frac{\partial{\cal{L}}}{\partial E} \ , \ \ \bar{\Pi} = \Pi - fr^2 \ , \ \ s = \mbox{sign}(r^2 \Delta + \bar{B}^2) \, .
\eequ
A $D2$-brane system with nonzero field strength can be thought of as
a bound state of $D2$-branes, $D0$-branes, and fundamental strings. 
The  conjugate momentum $N\Pi$ is just the 
number of strings that wrap $y$, and  $NB$ is the number
of $D0$-branes per unit length in the $y$ direction.
For configurations where $NB$ and $N\Pi$ are both positive, the  
system has a stationary solution which obeys the BPS conditions
\bequ
r_{BPS} = \sqrt{\Pi B} \ , \ \ {\cal{H}}_{BPS} = N\Pi + NB \, .
\eequ
As discussed in \cite{BHH}, the BPS condition, which usually gives a lower bound on ${\cal H}$, in some cases
gives an upper bound in the G\"odel background. When  $s = -1$, 
the kinetic term for the radial mode changes sign. The BPS solutions that exist with $s=-1$ actually sit at a
maximum of the effective potential\footnote{ The effective potential is defined as ${\cal H}$ in (\ref{Ham}) 
with $P_r$ set to zero.}
and are stable since the kinetic term is negative. We emphasize that stability here is
defined as stability in $t$ under radial perturbations of the form $r(t)$. As we will
discuss further in the next section, this may not always 
be the natural definition of stability
when there are closed timelike curves on the worldvolume.     

Notice that (\ref{Ham}) is finite at the VLS if $| N \bar{\Pi} \bar{B} | =  - N \bar{\Pi} \bar{B} $ and divergent
otherwise. As we will see later, the quantity $-N\bar{\Pi} \bar{B}$ is proportional to the angular momentum
of the probe. Probes with negative angular momentum (at the VLS) are unable to move through the VLS, while those
with positive angular momentum can freely pass, assuming this is allowed by kinematics.

It was shown in \cite{BHH} that the supertube probe computation is identical to a 
Nambu-Goto string probe computation on the background
U-dual to (\ref{back}), given by
\bequ
\barr{c}
ds^2=-dt^2+dy^2+2fr^2d\theta (dy-dt)+dr^2+r^2d\theta^2+\delta_{ij}dx^idx^j \ , 
\spa{0.3} \\
B_{NS}=- fr^2d\theta\wedge (dy-dt) \ , \earr 
\eequ 
where this $y$ is compact with period $2\pi R$. 
This is the compactified pp-wave, which is obtained from the Penrose limit of the NS five-brane near
horizon geometry. 
The identification of U-dual variables is as follows.  
\bequ
N \rightarrow -\omega^\prime \ , \ \
NB \rightarrow {R\omega} \ , \ \
N\Pi \rightarrow p_y \ , \ \
\eequ
where $p_y$ is the momentum in the $y$ direction, $w$ is the winding around the $y$ direction, and $w'$
is the non-topological winding around the $\theta$ direction. 
This is discussed further in Appendix B.

\sect{Time Traveling Supertube Probes}

The Hamiltonian (\ref{Ham}) adequately describes the 
classical motion of the probe in the interior of the VLS. Yet, in many cases
there appears to be a problem defining the Hamiltonian evolution outside the VLS. 
Typically, one finds that $\dot{r}$ is driven to infinity, at which point it is impossible
to continue the evolution. It turns out that in
these cases the choice of static gauge was inappropriate. Instead, taking $(\lambda , \xi_1 , \xi_2)$ to
be the worldvolume coordinates on the probe, we make the the following ansatz.
\begin{eqnarray}
t &=& t(\lambda) \nn \\ 
r &=& r(\lambda) \nn \\
y &=& \xi_1 \nn \\
\theta &=& \xi_2 \\
A_{\xi_1} &=& -B \xi_2 + A(\lambda)
\label{ansatz2}
\end{eqnarray}
with all the other spacetime coordinates taken to be constants and gauge fields set to zero.
Then, we find
\bequ
{\cal{L}} = -|N| \sqrt{ (-{\dot{r}}^2 +\dot{t}^2\Delta^{-1})( {r^2}\Delta + \bar{B}^2 ) - 
\bar{E}^2 {r^2}\Delta } 
-\dot{t} Nfr^2  + Nfr^2E \, ,
\label{Lag}
\eequ 
where
\bequ
\Delta = 1 -f^2 r^2 \ , \ \ 
\bar{B} = B - fr^2  \ , \ \
\bar{E} = E - {\dot{t} f\bar{B}}\Delta^{-1} \ ,
\eequ
and now the dot indicates differentiation with respect to $\lambda$, and $E=F_{\lambda \xi_1}$ and $B=F_{\xi_1 \xi_2}$
are the nonzero electromagnetic fields on the brane.
Our new Lagrangian can be derived using (\ref{gen}), or alternatively
by inserting $\dot{t}$ s into (\ref{olag}) wherever necessary to insure a reparametrization invariant action
whose Lagrangian matches with (\ref{olag}) when setting $t(\lambda) = \lambda$. 
We will think of the coordinate $\lambda$
as 
a worldvolume time coordinate and speak in those terms, 
but in fact as we will see shortly, the curve 
generated by $\frac{d}{d\lambda}$
on the worldvolume need not always be timelike. 
\bequ
\frac{d}{d\lambda} = \dot{t}\partial_t + \dot{r}\partial_r
\eequ 
However, even when this vector is spacelike the induced metric can remain Lorentzian.
\bequ
\sqrt{-\mbox{det}\, G} = \sqrt{ (-{\dot{r}}^2\Delta +\dot{t}^2) {r^2}}
\eequ
For example, when $\Delta<0$, $\dot{t}$ can vanish while the square root of the 
Lagrangian remains real.
Furthermore, even when $\frac{d}{d\lambda}$ is timelike, the worldvolume slices
of constant $\lambda$ may not be spacelike. One can ask whether or not
$\lambda$ is a \lq good' evolution parameter.
We will work only at the level of the equations of motion.
Solutions can be found in the usual way, 
by varying the fields in an action and considering the
resulting equations of motion, without
enforcing any particular boundary conditions and 
therefore without addressing a Cauchy problem. 
One can check that 
the ansatz (\ref{ansatz2}) is consistent with the full
equations of motion, so we will focus on the 
Lagrangian (\ref{Lag}) and simply look for solutions to
the equations of motion there. At this level, 
the fact that $\frac{d}{d\lambda}$ may not be timelike or
that $\lambda$ slicing does not define spacelike surfaces 
is irrelevant. 

On the other hand, it should be possible to define coordinates so that
the spacelike slices of the probe can be evolved through a
timelike coordinate. For these purposes, it is convenient
to consider the U-dual setup, which describes a string probe
on a compactified pp-wave. The string coordinates are taken to
be $\tau$ and $\sigma$, where $\sigma$ has period $2\pi$. The
supertube solutions that we will find can be carried over directly to the string
case with $\lambda \rightarrow \tau$. Additionally, the solutions
can be put into \lq light cone' gauge by a simple
reparametrization, so that the induced metric on the string is proportional to $\eta_{\alpha \beta}$.
In this gauge, it is a simple matter to define spacelike slices.
As shown in \cite{BHH}, the norm of the vector $\frac{d}{d \sigma}$ with
respect to the induced metric on the worldsheet is
proportional (with sign) to $s$. Thus, supertube configurations for which $s=-1$ are U-dual to strings
that wrap around closed timelike curves. Of course, if the string wraps around a closed
timelike curve, then $\sigma$ should be thought of as a worldsheet time coordinate. And
since the worldvolume is Lorentzian, $\tau$ should resemble\footnote{
To be precise, there is a reparametrization which is necesary before $\frac{d}{d \tau}$
has an everywhere nonzero norm.} a spacelike coordinate. 
Since $\tau$ is not periodic, a spacelike slice of the string will
have an infinite length, while a timelike slice will be  compact. 
We now have an explanation for the appearance
of the negative kinetic terms which occur when $s=-1$: They are not really kinetic terms
in a timelike evolution of a spacelike slice of the string. Rather, they are the gradient
terms which have the expected sign. 
This raises another question: What is the meaning of the stability of the probes with
$s=-1$? For the string, we could define the stability of the radial mode, for example, with
respect to the timelike $\sigma$ coordinate in the usual way 
by ignoring the periodicity of $\sigma$.
We will not carry out this calculation, but simply note that there may be more natural
definitions of the kinetic terms and of the stability of the BPS supertube probes which would follow from the U-dual
system.

It will be useful to work with the Routhian, ${\cal R} = {\cal L} - (N\Pi) E$, which is given by
\bequ
{\cal R} = \frac{ - s^{\prime} |N|}{ r^2  \Delta }\sqrt{ r^2  \Delta ( -\dot{r}^2 +
\dot{t}^2  \Delta^{-1} )( r^2 \Delta+\bar{\Pi}^2 ) 
(r^2 \Delta + \bar{B}^2) }- \dot{t} \frac{Nf}{\Delta}(r^2\Delta + \bar{\Pi}\bar{B}) \ ,
\eequ
and which serves as a Lagrangian for $r$ and $t$. The conjugate momentum to $E$ is
defined in the same way as the previous section, and
\bequ
s' = \mbox{sign}( r^2 \Delta+\bar{\Pi}^2) \ .
\eequ
Since there is no explicit $t$ dependence, the Routhian defines a conserved
quantity $H = -\frac{ \partial {\cal R} }{ \partial \dot{t}}$, which we will call the energy, given by
\bequ
H = \frac{ \dot{t}\, |N( r^2 \Delta+\bar{\Pi}^2) | 
(r^2 \Delta + \bar{B}^2)}{ \Delta{\left({ r^2  \Delta ( -\dot{r}^2 +
\dot{t}^2  \Delta^{-1} )( r^2 \Delta+\bar{\Pi}^2 ) 
(r^2 \Delta + \bar{B}^2) }\right)^{\frac{1}{2}} }}
+\frac{Nf}{\Delta}(r^2\Delta + \bar{\Pi}\bar{B}) \ ,
\label{H}
\eequ
which is equal to $\cal{H}$ in (\ref{Ham}) when $\dot{t}$ is set to one.
Notice that the system is invariant under
\bequ
\dot{t}\rightarrow -\dot{t} \, , \ \ \ N \rightarrow -N \, , \  \ \ H \rightarrow -H \, .
\eequ
Physically, this means that a probe 
traveling forward (backward) in time with energy $H$ can be 
interpreted as the charge conjugate\footnote{Flipping the sign of $N$ changes
the sign of the $D2$, $D0$, and $F1$ charges of the probe.} probe
($N \rightarrow -N$) traveling backward (forward) in time with energy $-H$. One should also
be aware of the odd nature of the square root in (\ref{H}). Although the quantity inside the
square root in manifestly positive within the VLS, in the regions where $( r^2 \Delta+\bar{\Pi}^2)>0$ and
$( r^2 \Delta+\bar{B}^2)<0$, or vice versa, the quantity is manifestly negative. Therefore, the probe is classically
forbidden to enter or exist in such regions. 

The equations of motion can be solved explicitly using (\ref{H}). The details of the solution can be found
in Appendix A. It turns out to be more convenient to solve the the equations of motion in terms of $x=r^2$. 
We will now briefly review the form of the probe trajectories. When 
\bequ
H = -f^{-1}N + N( \Pi +B) \equiv H_{\infty} \, ,
\eequ
which is the value of $H$ in (\ref{H}) when $x=\infty$, the solution is given by
\begin{eqnarray}
x(\lambda) &=& x_t \cosh^2(\lambda) \, , \nn \\
t(\lambda) &=&  \beta \sinh (\lambda) \, ,
\end{eqnarray}
where the coefficients $x_t$ and $\beta$ can be determined. This describes 
the probe in the far past contracting until it reaches the turning point $x_t$ at which point it begins to
expand and does so for the rest of its future. This turns out to be the  only case when the solution is 
unbounded in $x$. All other solutions oscillate between two radial turning points.
\begin{eqnarray}
x(\lambda) &=&  x_0 - \hat{x}  \sin{\lambda} \, , \nn \\
t(\lambda) &=& T_0 \lambda - \hat{T} \cos{\lambda} \ .
\label{solution}
\end{eqnarray}
The first equation describes radial oscillation about the midpoint $x_0$.
For $|T_0| > |\hat{T}|$, the second equation describes the probe traveling unidirectionally through time.
When $|T_0| < |\hat{T}|$ the probe travels both backwards and forwards in time along different parts of 
its trajectory. The net
drift\footnote{The solution in $x$ and $t$ is similar to the trajectory (in two spacelike directions) 
of a charged particle
in a background electromagnetic field with non-zero $E\times B$. The average velocity of the charged particle is
proportional to the vector $E\times B$, and the net motion is referred to as an $E\times B$ drift. }
through time is determined by the magnitude and sign of $T_0$. 
In the case when $T_0$ vanishes, (\ref{solution}) describes a periodic
geodesic.

\sect{Closed Geodesics}
The coefficients of the solution (\ref{solution}) are a bit complicated, but simplify 
in certain cases. When $\Pi = B$, 
they are given by
\begin{eqnarray}
x_0 &=& \frac{ (H - H_+)(H - H_-)}{ 2f^2( H - H_\infty)^2} \, \  , \nn \\
\hat{x} &=& \frac{ \sqrt{H^2(H-H_{BPS})(H - H_{non-BPS})}}{ 2f^2( H - H_\infty)^2} \, \ , \nn\\
T_0 &=& \frac{ s  H( H - 2H_\infty) }{4f( H - H_\infty)| H - H_\infty|} \ , \ \nn \\
\hat{T} &=& \frac{s \sqrt{ H^2(H-H_{BPS})(H-H_{non-BPS})} }{ 4f( H - H_\infty)| H - H_\infty|} \, \ ,
\label{coef2}
\end{eqnarray}
where
\begin{eqnarray}
H_{BPS} &=& 2NB \, \  , \nn \\
H_{non-BPS} &=& -2NB(1-2fB) \, \  , \nn \\
H_{\infty} &=& -f^{-1} N (1 - 2fB) \, \  , \nn \\
H_{\pm} &=& fNB^2 \pm \sqrt{N^2B^2[(1-fB)^2 +1-2fB]} \, \  .
\end{eqnarray}
We will further restrict attention to cases when $B<0 \, , \ N<0$, in which case $s=1$, and we have
the following hierarchy.
\bequ
H_{non-BPS} \ <  \ H_- \ < \ 0  \ <  \ H_+ \ < \ H_{BPS} \ < \ H_{\infty} \ < \ 2H_{\infty}
\eequ 
When $H \le H_{non-BPS}$, a solution exists, but $T_0 < 0$, so this should properly be interpreted
as the charge conjugate probe ($N \rightarrow -N \, , \ H \rightarrow -H$) traveling forward in time and
confined within the \lq potential'\footnote{Because the dynamics are described by a BI action, the \lq potentials' are only useful guides when the radial momentum is small. } well  plotted in Figure 1b. 
For $ H_{non-BPS} < H < H_{BPS}$ some of the coefficients (\ref{coef2}) become imaginary, indicating that no
solutions exist in this range. At $H = H_{BPS}$, we have a stationary solution. 
\begin{figure}
\begin{picture}(0,0)(0,0)
\put(-10,190){${\cal H}$}
\put(450,95){$x$}
\put(40,37){$x_0 = B^2$}
\put(340,95){$x_{VLS}$}
\put(160,40){$x$}
\put(245,230){${\cal H}$}
\end{picture}   
\centering\epsfig{file=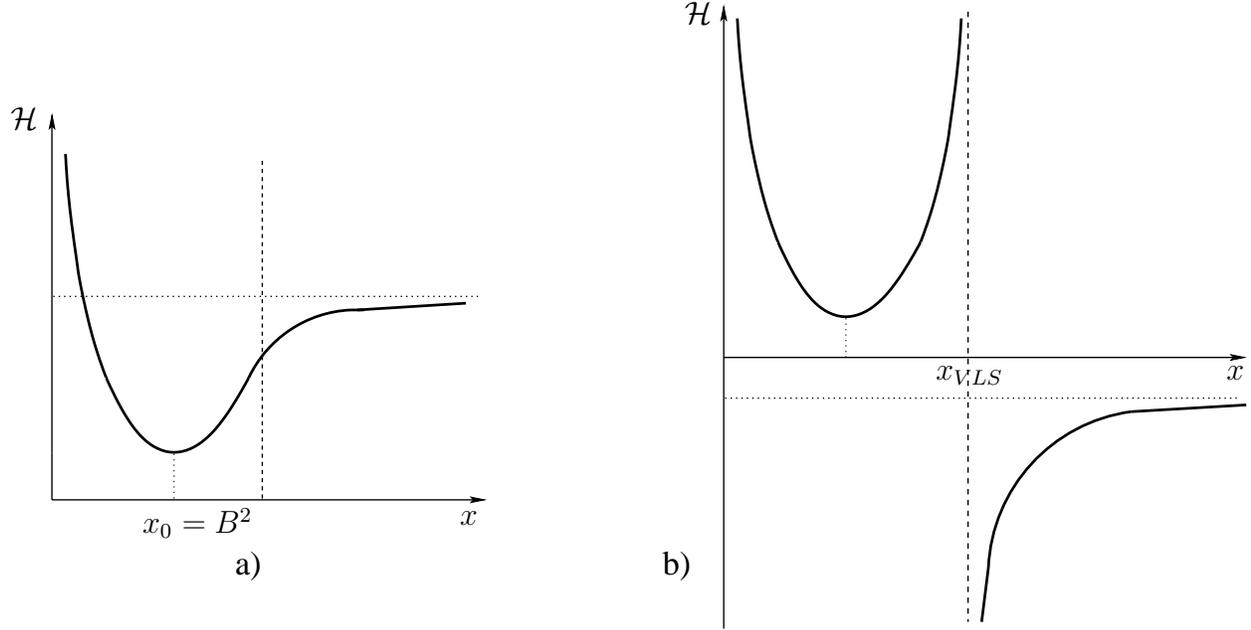,width=16cm}   
\caption{Effective potentials corresponding to $\Pi = B < 0$ with $N<0$ (left) and $N>0$ (right). The effective 
potential is defined as the Hamiltonian ${\cal H}$ given in (\ref{Ham}) with $P_r = 0$. }
\label{ctccre}
\end{figure} 
This is exactly when the
probe is at the minimum of the \lq potential'
plotted in Figure 1a, 
with $x_0 = B^2$. 
As we increase $H$ further, the solution shows oscillatory behavior in $x$. In the region $H_{BPS}<H<H_{\infty}$,
one can show that $T_0 > | \hat{T} |$, which implies the probe never travels backward in time. 
When $H = H_{\infty}$, the probe is no longer bound and
escapes to infinity. When $H_{\infty} <H < 2H_{\infty}$, the probe has a net drift backward in time, although
when it is within the VLS, it is always traveling forward in time. For $H>2H_{\infty}$, the drift returns to
being positive. Exactly when $H=2H_{\infty}$, the orbit is periodic. In this case the coefficients simplify
further.
\begin{eqnarray}
x_0 &=& \frac{ 2}{f^2} + \frac{B^2}{1-2fB}  \, \  , \nn \\
\hat{x} &=& \frac{2 (1-fB)}{f^2\sqrt{1-2fB}} \, \  , \nn \\
T_0 &=& 0  \, \  , \nn \\ 
\hat{T} &=&  \frac{ (1-fB)}{f\sqrt{1-2fB}} \, \  .
\label{closedcoef}
\end{eqnarray}
From these expressions we see that we can relax some of the constraints we imposed earlier. 
This is a valid solution
whenever $B<1/(2f)$ and $B\ne0$.
Although we have looked at a special case, by looking at the full solution in Appendix A, one can see 
that even when $\Pi \ne B$ periodic orbits are possible. 
In fact, for positive $H$ and non-vanishing $\Pi$ and $B$ whenever
$N<0$, $B<1/(2f)$ and  $\Pi<1/(2f)$, periodic geodesics exist when
\bequ
H= H_{\infty} - Nf^{-1}\sqrt{(1-2f\Pi)(1-2fB)} \ .
\eequ

In order to determine whether or not these periodic geodesics can be closed, we must determine 
the period of the gauge field $A(\lambda)$ as defined in (\ref{ansatz2}). 
Using a result from Appendix A, we have
\bequ
A(2\pi m) - A(0) =  \int^{2\pi m}_0 d\lambda E(\lambda) = -\frac{2\pi m}{2f}\left( 1 + \sqrt{\frac{1-2fB}{1-2f\Pi}} \right) \ ,
\eequ
which is a valid expression even when $\Pi \ne B$. 
Here, $m$ is an arbitrary integer.
Clearly this would imply that $A(\lambda)$ is not periodic, but we have yet
to take into account gauge equivalence. A gauge transformation by the element $g=$exp$(\frac{iyn}{L})$, which is
single valued\footnote{ For $|N|>1$ there may be gauge transformations by 
group elements that are not single valued, but rather are sections
on a twisted bundle. These can result in \lq tighter' identifications such as $A \sim A + \frac{1}{NL}$.}
on the worldvolume if $n$ is an integer, produces the following identification.
\bequ
A \sim A + \frac{n}{ L} \ .
\eequ
Thus if\footnote{In this expression, we have reintroduced an appropriate factor of $2\pi \alpha'$.}
\bequ
\frac{L}{2f\alpha'}\left( 1 + \sqrt{\frac{1-2fB}{1-2f\Pi}} \right) = \frac{n}{m}
\label{rash}
\eequ
for some integers $n$ and $m$ - or in other words, if the left hand side of the above equation is rational - 
then $A(\lambda)$ is periodic in $\lambda$ with period $2 \pi m$. 
In this case,
the affine parameter can be identified 
\bequ
\lambda \sim \lambda + 2\pi m 
\eequ
and the geodesic is closed. Since $\Pi$ can be varied\footnote{At the classical level $\Pi$ is not
quantized, whereas $B$ is. However, if we assume the geodesic is closed, 
(\ref{rash}) can be considered a quantization condition on $\Pi$, in
the same way $B$ is quantized. That is, $\int_{\cal C_i} F = 2\pi n^i$, 
where the integral is over any worldvolume 2-cycle ${\cal C}_i$ and the $n^i$ are integers.} continuously, for a fixed background  and magnetic field, 
$\Pi$ can always be adjusted so that the
above quantity is rational.
\sect{Gravitational Couplings}

The probe's contribution to the energy momentum tensor can be calculated.
\bequ
T^{\mu \nu} = \frac{-2}{\sqrt{g}}  \frac{ \delta S }{ \delta g_{\mu \nu}} \ .
\eequ
It is useful to define an energy momentum density ${\cal T}$ on the probe through the expression
\bequ
\sqrt{g} \ T^{\mu}_{\, \ \nu} = \int d^2\xi \, d\lambda \ {\cal T}^{\mu}_{\, \ \nu} \ \delta^{10}(X - X(\xi,\lambda)) \ ,
\eequ
where $X$ represents all the spacetime coordinates and $\lambda$ and $\xi$ parameterize the spacetime embedding of the
probe.
Then using the Lagrangian (\ref{gen}), we find
\begin{eqnarray}
{\cal T}^0_{\, \ 0} &=& \frac{ \dot{t}^2\, |N( r^2 \Delta+\bar{\Pi}^2) |   
(r^2 \Delta + \bar{B}^2)}{ \Delta{\left({ r^2  \Delta ( -\dot{r}^2 +
\dot{t}^2  \Delta^{-1} )( r^2 \Delta+\bar{\Pi}^2 ) 
(r^2 \Delta + \bar{B}^2) }\right)^{\frac{1}{2}} }}
+\frac{\dot{t} Nf \, \bar{\Pi}\bar{B}}{\Delta} \nn \\
&=& \dot{t}(H - Nfr^2) \ , \\
{\cal T}^0_{\, \ \theta} &=& \dot{t}(-N\bar{\Pi}\bar{B}) \ .
\end{eqnarray}
In these expressions we have already fixed the $\xi$ parametrization, as we had done in the
Lagrangian (\ref{Lag}).

Let us consider a supertube traveling along a trajectory corresponding to a periodic geodesic. 
For the trajectories we found with $\Pi=B$ both $(H - Nfr^2)>0$ and $(-N\bar{\Pi}\bar{B}) > 0$, indicating 
that ${T}^0_{\, \ 0}$ and ${T}^0_{\, \ \theta}$ are positive when the probe is traveling
forward in time (closer to the origin) and negative when traveling backward in time.  Now consider the
following integral,
\bequ
\int_{{\cal N}}\sqrt{g} \ T^{0}_{\, \ 0} = \left(\int d^2\xi \right) \int d\lambda \, 
{\cal T}^0_{\, \ 0} \ ,
\eequ
where the first integral is over some bounded region ${\cal N}$ of {\it finite} spacetime volume 
containing the entire periodic geodesic. The double integral
over the two parameters $\xi$ gives the constant $(2 \pi)^2 L$.
The
integration over the last affine parameter $\lambda$ will be divergent unless
$\lambda$ can be periodically identified. As discussed in the last section, this is possible for the case when $\Pi=B$ 
if $f\alpha' /L$ is rational. If this quantity is irrational or if the
periodic identification is not made, then the above integral will
diverge as a result of the unbounded integration over $\lambda$, and the probe approximation is not valid. 
The same result holds for  $T^{0}_{\, \ \theta}$, and we will have similar
results for all the other nonzero components of the energy momentum tensor.

One can also study the effects of the closed supertube geodesics 
in the framework suggested in \cite{Drukker:2003sc}
where a supertube domain wall sources a background which in some bounded region reproduces 
the G\"odel background studied here. 
In this construction, 
the region where $g_{\theta \theta}<0$ is bounded within 
some shell and the timelike curves that do not enter this region cannot be closed. This G\"odel background is U-dual to a compactified pp-wave,
and the string theory on this background can be quantized \cite{Russo:1994cv}. 
In Appendix B, we translate the supertube
probe computation to a string probe calculation on this U-dual background. This seems like the
natural place to flush out the effects  of these closed geodesics at the quantum level, but we leave this
for future work.
We should also note that the supertubes likely decouple when the compactification
radius $L$ is taken to infinity, while in the G\"odel background (\ref{back}), the closed timelike curves continue
to exist. Thus, if one is of the opinion that
the closed D2 brane geodesics signal problems for string 
theory on this background, 
one still has the decompactification limit to consider. In the noncompact case, there may
be suitable holographic screens \cite{Boyda:2002ba, Brecher:2003rv}, which would ensure chronology protection.

\section *{Acknowledgments}
We would like to thank Jason Fleisher and Vika Naipak for assistance. We are especially grateful to Oren Bergman,
Nadav Drukker, and Shinji Hirano for useful discussions.
This work was supported by the Israel Science Foundation under grant No. 101/01-1. 

\setcounter{section}{1}
\renewcommand{\theequation}{\Alph{section}.\arabic{equation}}
\section *{Appendix A}
\setcounter{equation}{0}

Before solving the differential equation (\ref{H}), we will first demonstrate the method of
solution on a simpler example.
\subsection *{The Harmonic Oscillator}
Consider the Lagrangian for a harmonic oscillator
\bequ
L = \frac{1}{2}\, \dot{x}^2 - \frac{1}{2} \, \omega^2 x^2 \ .
\eequ
The Hamiltonian ${H} = \dot{x} p - L$, can be obtained in the usual way
\bequ
{{H}} = \frac{1}{2} {p^2} +  \frac{1}{2} \omega^2 x^2 \ ,
\eequ
and the equations of motion can be solved. However, we would like to find the harmonic oscillator solutions
using a different technique.

We begin by rewriting the Lagrangian in a such a way that the action is reparametrization invariant.
\bequ
L = \frac{1}{2 \dot{t}} \, \dot{x}^2 - \frac{\dot{t}}{2} \, \omega^2 x^2 \ ,
\eequ
where now the dot indicates differentiation with respect to the affine parameter $\lambda$. 
Setting $\lambda=t$, we recover the original Lagrangian.
Since the Lagrangian contains no explicit $t$ dependence, there is a conserved quantity
\bequ
H = - \frac{ \partial L}{ \partial \dot{t}} = \frac{1}{2 \dot{t}^2} \, \dot{x}^2 + \frac{1}{2} \, \omega^2 x^2 \ .
\eequ
In order to solve for the motion, we first solve for $\dot{t}$.
We find
\bequ
\dot{t} = \frac{\pm | \dot{x} \,  |}{ \sqrt{2H - \omega^2 x^2}} = \frac{\pm | \dot{x} \,  |}{ \sqrt{p^2}} \ .
\label{poo}
\eequ
Notice that $ p^2 \equiv 2H- \omega^2 x^2 $ must be greater than or equal to zero for physical solutions.
The turning points occur when $p^2 = 0$, or equivalently when $ x = \pm \sqrt{2H}/\omega$. 
We can fix the reparametrization invariance by
setting,
\bequ
x = \frac{\sqrt{2H}}{\omega} \, \sin (\lambda) \ .
\eequ
Plugging this back into (\ref{poo}), we find $\dot{t} = \pm \omega^{-1}$, and thus $\lambda = \pm \omega ( t - t_0)$.
We then arrive at the familiar result
\bequ
x = \frac{\sqrt{2H}}{\omega} \, \sin (\pm \omega ( t- t_0) ) \ .
\eequ
Notice, in this example $\dot{t}$ can never vanish, so the plus and minus 
signs are simply related by the time reversal
symmetry of the original action.

\subsection *{The Supertube}
In order to solve for the supertube motion, we can solve (\ref{H}) for $\dot{t}$. We find
\bequ
\dot{t} = \frac{ s \bar{H} \Delta r |\dot{r}\, |}{ \left(r^2\Delta \bar{H}^2 - \Delta^{-1}N^2 ( r^2 \Delta+\bar{\Pi}^2 )
(r^2 \Delta + \bar{B}^2)\right)^{\frac{1}{2}}}
\label{tdot}
\eequ
where
\bequ
\bar{H} = H -\frac{Nf}{\Delta}(r^2\Delta + \bar{\Pi}\bar{B}) \ .
\eequ
At this point, it is convenient to use the variable $x=r^2$. Then one can show that the object inside the 
square root is proportional to the square of the momentum conjugate to $x$,
in analogy with the harmonic oscillator example.
Let us rewrite everything in terms of $x$.
\begin{eqnarray}
\dot{t} &=& \frac{s \bar{H} \Delta | \dot{x} \, |}{ 2 \sqrt{ P_x^2}} \label{tdot2}  \ , \nn \\
P_x^2 &\equiv& -(N \Pi B)^2 + x \left[ ( H - f N \Pi B )^2 - N^2( \Pi + B - f \Pi B)^2 +2N^2\Pi B \right]   \nn \\
&&-x^2\left[Hf +N -fN( \Pi +B)\right]^2  \ , \nn \\
\bar{H} \Delta &=& (H-fN\Pi B) - x f\left[Hf +N - fN( \Pi + B)\right] \ . 
\label{stuff}
\end{eqnarray}
%
The zeros of $P_x^2$ determine the radial position of the turning points where $\dot{x}$ is zero. 
Similarly, the zero of $\bar{H} \Delta$ determines where $\dot{t}$ is zero.
Since $P_x^2$ must be positive, this leads to bounds on the physical values of $H$.
Also, since $P_x^2$ is generically quadratic in $x$, it has at most two zeros. When the two
zeros are identical, this corresponds to a static configuration with vanishing radial momentum. When
the two zeros are distinct, the radius oscillates between them. For nonzero $\Pi$ and $B$, $P_x^2$ is negative
at the origin. Therefore, if $P_x^2$ has no zeros, $P_x^2$ is always negative, 
and this corresponds to an unphysical value
of $H$. We can conclude that all generic solutions with nonzero  $\Pi$ and $B$ perform oscillatory motion. For
special values of $H$, the quadratic $P_x^2$ may reduce to a linear, in which case there is only one turning point. 
By considering (\ref{tdot}) one can also see that  $P_x^2$ is manifestly negative in the region where 
$(r^2 \Delta + \bar{B}^2)> 0$ and $(r^2 \Delta + \bar{\Pi}^2)< 0$, or
vice versa. Therefore, both zeros occur either before or both after this physically unallowed region.

After fixing some $H$, one can solve for the zeros of $P_x^2$. 
Let the zeros be given by $x_1$ and
$x_2$. Then defining
\bequ
x(\lambda) = \frac{(x_1 + x_2)}{2} - \frac{(x_2 - x_1)}{2} \sin (\lambda) \ ,  
\eequ
and plugging back into (\ref{tdot2}), $\dot{t}$ can be integrated. Of course, if there is only one turning point
then we should define
\bequ
x(\lambda) = x_1 \cosh^2 (\lambda) \, .  
\label{cosh}
\eequ
It is convenient to make some further definitions.
\bequ
P_x^2 = -C^2 + bx -A^2x^2 \, , \ \ \   \bar{H}\Delta = d - xfA \ ,
\eequ
where the new coefficients can be read off from (\ref{stuff}).
After setting
\bequ
x = \frac{ b}{2A^2} - \frac{ \sqrt{b^2 - 4A^2C^2}}{2A^2} \sin{\lambda} \ ,
\eequ
we find
\bequ
\frac{| \dot{x} \, |}{ \sqrt{P_x^2}} = |A|^{-1} \ .
\eequ
Plugging this result into (\ref{tdot2}), we have
\bequ
\dot{t} = \frac{s}{4 A |A| } \left( (2Ad - fb) + f\sqrt{b^2 - 4A^2C^2}  \ \sin{\lambda}\right) \ ,
\eequ
which can be integrated to give
\bequ
{t} - t_0 = \frac{s}{4 A |A\,| } \left( (2Ad - fb) \lambda - f\sqrt{b^2 - 4A^2C^2}  \ \cos{\lambda}\right) \ .
\eequ
Notice in the limit $f\rightarrow 0$, we find $t = H\lambda / 2 |N \,|$, which gives the frequency of
supertube radial oscillation in flat space, $\omega = 2 | N \, | / H$.
%
%
Ignoring integration constants, we write the final form of the solution,
\begin{eqnarray}
x(\lambda) &=&  x_0 - \hat{x}  \sin{\lambda}  \ , \nn\\
t(\lambda) &=& T_0 \lambda - \hat{T} \cos{\lambda} \ ,
\end{eqnarray}
where the coefficients are given by
\begin{eqnarray}
x_0 &=& \frac{ (H - H_{4+})(H - H_{4-})}{ 2f^2( H - H_\infty)^2}  \ , \nn \\
\hat{x} &=& \frac{ \sqrt{(H-H_{BPS})(H - H_1)(H - H_2)(H - H_3)}}{ 2f^2( H - H_\infty)^2}  \ , \nn\\
T_0 &=& \frac{ s  (H - H_{5+})( H - H_{5-}) }{4f( H - H_\infty)| H - H_\infty|}  \ , \nn\\
\hat{T} &=& \frac{s \sqrt{(H-H_{BPS})(H - H_1)(H - H_2)(H - H_3) } }{ 4f( H - H_\infty)| H - H_\infty|} \ , 
\label{coef}
\end{eqnarray}
and where
\begin{eqnarray}
H_{BPS} &=& N(\Pi+B) \ , \nn \\
H_{\infty} &=& -f^{-1} N  + N(\Pi + B) \ , \nn \\
H_{4\pm} &=& fN\Pi B \pm \sqrt{N^2(\Pi + B -f\Pi B)^2 -2N^2\Pi B} \ , \nn \\
H_{5\pm} &=& H_{\infty} \pm \sqrt{ H_{\infty}^2 - N^2(\Pi -B)^2} \ , \\
&=&  H_{\infty} \pm |N|f^{-1}\sqrt{(1-2f\Pi)(1-2fB)} \ .
\end{eqnarray}
The yet undefined quantities are given by the roots of the quartic.
\bequ
b^2 - 4A^2C^2 = (H - H_{BPS})(H - H_1)(H -H_2)(H-H_3).
\eequ 

We will now derive an expression for the electric field $E$, which will be useful for understanding
when a periodic geodesic can be closed.
From the definition of the conjugate momentum $\Pi$, we have
\bequ
E = \frac{n \, \bar{\Pi}}{r^2 \Delta |  r^2 \Delta +\bar{\Pi}^2 |} \left(r^2 \Delta (-\dot{r}^2 + \dot{t}^2
\Delta^{-1})  ( r^2 \Delta +\bar{\Pi}^2 ) ( r^2 \Delta +\bar{B}^2 )
\right)^\frac{1}{2}  + \frac{\dot{t} f \bar{B} }{\Delta} \ ,
\eequ
where $n = \mbox{sign}(N)$. From this expression, we see that $E$ diverges at the VLS if the probes angular
momentum is negative, just as the Hamiltonian does. 
Although the parametrization we chose was simple for $x$ and $t$, it does not come out so nicely for $E$.
\bequ
E = \frac{sB ( N \Pi B + x  f(H - H_{BPS}))}{2fx | H-H_\infty|} +
\frac{s( N\Pi - fN\Pi B -xf^2 (H-H_\infty))}{2f | H-H_\infty|} \ ,
\label{Ef}
\eequ
where one must plug in the solution $x(\lambda)$. 
One can check that when $H=H_{BPS}$ with $x=\Pi B$, the electric field satisfies $E/\dot{t} =1$, which
is the expected result for the BPS solution.

When the geodesics are periodic, $T_0$ vanishes. For configurations where $N<0$, this determines that $H=H_{5+}$.
Plugging this value of $H$ into (\ref{Ef}), we find 
\bequ
2\pi Y \equiv \int^{2\pi}_0 d\lambda E(\lambda) = -\frac{2\pi}{2f}\left( 1 + \sqrt{\frac{1-2fB}{1-2f\Pi}} \right) 
\ . \ \ \ \ \ \ \ (\mbox{periodic geodesics})
\label{intE}
\eequ
Before obtaining this expression, we made use of the following intermediate result for $x(\lambda)$ when $H=H_{5+}$.
\begin{eqnarray}
\hat{x} &=& \frac{\sqrt{1-2f\Pi}(1-fB) + \sqrt{1-2fB}(1-f\Pi) }{f^2 \sqrt{(1-2f\Pi)(1-2fB)}} \nn \\
x_0 &=& \frac{1}{f^2}\left( 1 + \frac{ (1-f\Pi)(1-fB)}{\sqrt{(1-2f\Pi)(1-2fB)}} \right) \ .\ \ \ \ \ \ (\mbox{periodic geodesics})
\end {eqnarray}
The gauge field $A_{y}$ is periodically identified
\bequ
A_{y} \sim A_{y} + \frac{1}{ L}
\eequ
as can be seen by considering the $U(1)$ gauge transformation $g=\mbox{exp}(\frac{iy}{ L})$,
which is a single valued group element. The gauge field and the spacetime coordinates are 
then periodic in $\lambda$ if $YL/\alpha'$ is rational, where we reinserted an appropriate factor of $2\pi \alpha'$.

It will be useful to define a new quantity\footnote{ $p_+$ corresponds to the \lq light-cone' momentum
in the Polyakov string on the U-dual background, which we will discuss further in the next appendix.} $p_+$,
\bequ
p_+ = -H + N(\Pi + B) \ .
\eequ
For the case of periodic geodesics, we again set $H=H_{5+}$ and find
\bequ
p_+ = Nf^{-1} + Nf^{-1}\sqrt{(1-2f\Pi)(1-2fB)}      \ \ \ \ \ \ \ \ (\mbox{periodic geodesics})
\label{Pplus}
\eequ

\addtocounter{section}{1}
\setcounter{equation}{0}
\section *{Appendix B}
As shown in \cite{BHH}, the D2 brane probe calculation is identical to a string probe computation on the
U-dual background, which is the compactified pp-wave
\bequ
\barr{c}
ds^2=-dt^2+dy^2+2fr^2d\theta (dy-dt)+dr^2+r^2d\theta^2+\delta_{ij}dx^idx^j \ , \spa{.3} \\
B_{NS}=- fr^2d\theta\wedge (dy-dt) \ , \earr \label{ourst} 
\eequ 
where the $y$-direction is compactified with radius $R$. The dynamics of the string are described by the
Nambu-Goto Lagrangian.
\begin{equation}
{\cal L}=-\sqrt{-\det g} - B_{NS} ,
\label{NambuGoto}
\end{equation}
where $g$ and $B_{NS}$ are the induced metric and NS two form on the worldsheet.
The U-dual of the supertube solutions are described by the following ansatz\footnote{By a reparametrization of the
form $\sigma \rightarrow \sigma + h(\lambda)$, we can intoduce $\lambda$ dependence into the $\theta(\lambda, \tau)$ ansatz, while
at the same time shifting $y(\lambda) \rightarrow y(\lambda) + Rwh(\lambda)$. To put the string in \lq light cone' gauge, we can use this freedom to shift $y(\lambda)$ so that the first term in the untranslated solution for $dy/d\lambda$ (\ref{Ef}) is removed.
After a simple rescaling of $\lambda$, the string will be in \lq light cone' gauge.}
\begin{eqnarray}
t(\lambda,\sigma)&=&t(\lambda) \ ,\nn \\
y(\lambda,\sigma)&=&Rw\sigma+y(\lambda) \ ,\nn \\
\theta(\lambda,\sigma)&=&w'\sigma \ , \nn \\
r^2(\lambda,\sigma)&=&x(\lambda)\ ,
\label{ansatz}
\end{eqnarray}
where $\sigma$ has period $2\pi$. 
The ansatz describes a string centered at $r=0$ with winding $w$ around the compact $y$ direction and
non-topological winding $w'$ around the $\theta$ direction. The rest of the spacetime coordinates are taken to be 
constants.
Both $t(\lambda)$
and $x(\lambda)$ can be carried over directly from the supertube solution with the following substitutions.
\bequ
N \rightarrow -\omega^\prime \ , \ \
NB \rightarrow {R\omega} \ , \ \
N\Pi \rightarrow p_y \ , \ \
\eequ
where $p_y$ is the momentum conjugate to $y(\lambda)$. To get an explicit expression for $y(\lambda)$, we
just need to integrate 
\bequ
y(\lambda) = y_0 + \int^\lambda d\lambda '  \, E(\lambda ') \ ,
\eequ
and make the same substitutions. Here,  $E$ is given by (\ref{Ef}). When the U-dual supertube
geodesics are periodic, the string geodesics generically will not be since $y$ may not be periodic. 
In particular, translating the result (\ref{intE}) we find
\bequ
Y \equiv \frac{y(2\pi)-y(0)}{2\pi} = -\frac{1}{2f}\left( 1 + \sqrt{   \frac{1-2f \left(\frac{-Rw}{w'}\right)  }
{ 1-2f\left( \frac{-p_y}{w'} \right) } } \right) \ .
\eequ
When $Y/R$ is rational, the geodesic can be closed, otherwise
the two dimensional geodesic is dense on some three dimensional surface. For example, when $p_y = Rw$ ($\Pi=B$),
the string geodesics close if $fR$ is rational.

The string theory on this background has been 
quantized in \lq light cone' gauge \cite{Russo:1994cv}, where the
\lq light cone' momentum is 
given by\footnote{In the Polyokov framework, $p_+ = E + p_y +Rw$, but the energy $E$
from the Polyakov formalism should be identified with $-H$ from the Nambu Goto action. Also, the $f$ defined in
 \cite{Russo:1994cv} differs from the definition of $f$ in this paper by a factor of 2.}
\bequ
p_+ =  -H + p_y + Rw \ .
\eequ  
When $p_+$ is outside of the range $-1<fp_+<1$, the quantum string states belong to spectral flowed representations
of the Heisenberg current algebra. The use of  these representations is required to 
avoid the appearance of states with
negative norm.
For the problematic geodesics we have
\bequ
p_+ =  -w'f^{-1}  -w'f^{-1}\sqrt{\left(1-2f\left(\frac{p_y}{-w'}\right)\right)\left(1-2f\left(\frac{Rw}{-w'}\right) \right)} \ ,
\eequ
where we translated the result (\ref{Pplus}). Since $w'$ is a positive integer, $fp_+ < -1$, which implies the 
quantum states, out of which a corresponding coherent state might be constructed, 
should be found in spectral flowed representations.

\end{document}